\documentstyle[12pt]{article}
\title{\bf Nonperturbative contribution to the
Dokshitzer-Gribov-Lipatov-Altarelli-Parisi and Gribov-Levin-Ryskin equation}
\author{\bf Bo He\thanks{ Permanent Address: Department of Applied Physics,
Shanghai Jiaotong University, Minhang Campus,
Shanghai 200240, P.R. China. E-mail: hebobo@public6.sta.net.cn }\\
CCAST (World Laboratory), P.O.Box 8730 \\
Beijing 100080, P.R. China \\
Department of Applied Physics, Shanghai Jiaotong University\\
Shanghai 200030, P.R. China}

\date{}
\begin{document}
\maketitle

\begin{abstract}
By studying the nonperturbative contribution to the
Dokshitzer-Gribov-Lipatov-Altarelli-Parisi and Gribov-Levin-Ryskin
equation, it is found that (i) the nonperturbative contribution suppresses
the evolution rate at the low $Q^2$, small-x region; (ii) the
nonperturbative contribution weakens the shadowing effect. The method in this
paper suggests a smooth transition from the low $Q^2$ ("soft" ), where
nonperturbative contribution dominates, to the large $Q^2$ ("hard" ) region,
where the perturbative contribution dominates and  the nonperturbative
contribution can be neglected.

{\bf PACS} numbers:12.38.Aw, 13.60.Hb
\end{abstract}

\newpage

The properties of parton distribution at small-x region ( x is the value of
the Bjorken variable ) have recently
been the important subject [1-5]. Recent measurements of the structure
functions for the deep-inelastic {\it ep} scattering at HERA discovered
their dramatic rise as x decreases from $10^{-2}$ to $10^{-4}$ [6,7]. The predictions
of Gl\"{u}ck, Reya and Vogt (GRV) [8] by using the
Dokshitzer-Gribov-Lipatov-Altarelli-Parisi ( DGLAP ) evolution equation [9]
at very low $Q^2$ ( $Q^2$ is the negative of the square of the four-momentum
transferred by the lepton to the nucleon ) are in broad agreement
with this result. However, the GRV model
fails in low $Q^2$ quantitatively, that is to say, the evolution rate is faster
than that of the experiments. \footnote[1]{ Ref. [7] shows that the value of $F_2$
given by GRV model is lower than that of the experiments at $Q^2=0.4GeV^2$
while it is higher at $Q^2=6.5GeV^2$.}  It then becomes a challenging problem
how to determine the structure function at the low $Q^2$ region.
Another important question is whether the shadowing effect, which  the
Gribov-Levin-Ryskin (GLR) equation [10] describes,  can be observed by
the current experiments at HERA.

The purpose of this letter is to study the DGLAP and GLR
equation in low $Q^2$ by considering the nonperturbative contribution. It will
be showed that (i) the nonperturbative contribution suppresses
the evolution rate at the low $Q^2$, small-x region; (ii) the
nonperturbative contribution weakens the shadowing effect. So the
nonperturbative contribution is very important in the low $Q^2$ region.\\

The DGLAP equation for the gluon distribution at small-x region
in the DLLA is given by

 \begin{equation}\frac{\partial^{2} xg(x,Q^{2})}
{\partial \ln (1/x)\partial\ln (Q^{2})}=
\frac{3\alpha_{s}(Q^2)}{\pi}xg(x,Q^{2})\end{equation}

By considering the shadowing effect, the DGLAP equation can be modified
in the form ( called GLR equation):

\begin{equation}
\frac{\partial^{2} xg(x,Q^{2})}
{\partial \ln (1/x)\partial\ln (Q^{2})}=
\frac{3\alpha_{s}(Q^2)}{\pi}xg(x,Q^{2})-\frac{81\alpha_{s}^{2}(Q^2)}{16R^{2}Q^{2}}
[xg(x,Q^{2})]^{2}
\end{equation} \\

In the previous studies people
used the perturbative QCD effective coupling ( the leading order coupling)
$\alpha_{s}(Q^2)$  to study the equation, where
\begin{equation}\alpha_{s}(Q^2)=
\frac{12\pi}{(33-2n_{f})\log(Q^{2}/\Lambda^{2}_{QCD})}\end{equation}

By using the formula (3), the Eq.(2) can be cast in the form:
 \begin{equation}\partial_{y}\partial_{t}G(y,t)=cG(y,t)
 -\lambda\exp[-t-\exp(t)]G^{2}(y,t)\end{equation}
 where $y=\ln(1/x),t=\ln[\ln(Q^{2}/\Lambda^{2}_{QCD})]$,  $G(y,t)=xg(x,Q^{2})$,
 $c=12/(11-2n_{f}/3)$ with $n_f$ the number of quark flavors and
$\lambda=9\pi^{2}c^{2}/16R^{2}\Lambda^{2}_{QCD}$.
The equation (4) has been studied by many authors [11,12].
Eskola, Qiu and Wang [13]
applied the eq.(4) to study the shadowing in heavy nuclei.\\

However, it should be noted that in the large $Q^2$ the formula (3) derived
from perturbative QCD is a good approximation while in the low $Q^2$
nonperturbative contribution should be included. To avoid the ghost-pole
problem in the behavior of a running coupling, Shirkov and Solovtsov
[14] obtained the QCD running coupling in the leading order as:\\
\begin{equation}\alpha_{an}(Q^2)=
\frac{12\pi}{(33-2n_{f})}[\frac{1}{\log(Q^{2}/\Lambda^{2}_{QCD})}+\frac{1}
{1-Q^2/\Lambda^{2}_{QCD}}]
\end{equation}

The second term on the right-hand side of Eq. (5) is clearly the
nonperturbative contribution. It is noted that
$\alpha_{an}(Q^2=0GeV^2)=\frac{12\pi}{(33-2n_{f})}$
depends only on group factors and does not depend on the $\Lambda_{QCD}$.
It can be found that in the large $Q^2$ the running coupling $\alpha_{an}(Q^2)$
is dominated by perturbative contribution and  the nonperturbative
contribution can be neglected while in the low $Q^2$ the
nonperturbative contribution is very notable.\\

By applying the formula (5) to DGLAP equation \footnote[2]{The reason that
the formula (5) replaces formula(3) in DGLAP equation will be given in the
ending of the text.}, the Eq. (1) can be written as:
 \begin{equation}\partial_{y}\partial_{t}G(y,t)=cG(y,t)[1+\frac{\exp(t)}
 {1-\exp(e^t)}]
 \end{equation}

Adopting the semi-classical approximation [4,12], which amounts to keeping
only the first order derivatives of the function $z(y,t)=\log[G(y,t)]$,
Eq.(6) is rewritten as:
\begin{equation}\partial_{y}z(y,t)\partial_{t}z(y,t)=c[1+\frac{\exp(t)}
{1-\exp(e^t)}]
 \end{equation}

Eq. (7) can be solved by using the method of characteristic [4,12]. Let
\begin{equation}p=\partial_{t}z(y,t), \hspace{1.0cm}
 q=\partial_{y}z(y,t)  \end{equation}
Eq. (7) can be written as the following general form:
 \begin{equation}F(p,q,t,y,z)=0\end{equation}
 The characteristic differential equations of Eq. (9) have the following form:
\begin{eqnarray}
&&\frac{dt(\tau)}{d\tau}=F_{p},\hspace{1cm}
\frac{dy(\tau)}{d\tau}=F_{q},\hspace{1cm}
\frac{dz(\tau)}{d\tau}=pF_{p}+qF_{q},\nonumber \\
&&\frac{dp(\tau)}{d\tau}=-(F_{t}+pF_{z}),\hspace{1cm}
\frac{dq(\tau)}{d\tau}=-(F_{y}+qF_{z}),
\end{eqnarray}
where $\tau$ is the "inner" time.\\

$F_{p}, F_{q}, F_{t}, F_{y}, F_{z}$ is
\begin{eqnarray}
&&F_{p}=q,\hspace{1cm} F_{q}=p,\hspace{1cm}F_{y}=0,
\hspace{1cm}F_{z}=0. \nonumber \\
&&F_{t}=-ce^t[1-\exp(e^t)+\exp(e^t+t)]/[1-\exp(e^t)]^2
\end{eqnarray}

Same as Ref. [12], the initial conditions to solve Eqs. (10) are:
\begin{eqnarray}
&&t_{0}=\log[\log(Q^{2}_{0}/\Lambda^{2}_{QCD})]
=\log[\log(4GeV^{2}/\Lambda^{2}_{QCD})], \nonumber \\
&&y_{0}=\log(1/x_{0})=\log(100), p_{0}=c/\delta_{bare},
z_{0}=\log(3.38), \nonumber \\
&&q_0=c[1+\frac{e^{t_{0}}}{1-\exp(e^{t_{0}})}]/p_0, \Lambda_{QCD}=0.2GeV,
\delta_{bare}=0.5.
\end{eqnarray} \\

For clear comparison between the results of Eqs. (10) with those of
DGLAP equation not including the nonperturbative contribution , we
solve the Eqs. (10) numerically by adopting the Runge-Kutta
methods [12]. Fig. 1 shows the evolution path (y,t)
corresponding to the dashed line by solving the Eqs. (10) compared
with the evolution path (y,t) corresponding to the solid line by solving
the characteristic differential equations of DGLAP not including
nonperturbative contribution.\\

Like the DGLAP evolution equation, by using the formula (5),
the GLR equation can be cast in the form:
 \begin{equation}
 \partial_{y}\partial_{t}G(y,t)=cG(y,t)[1+\frac{\exp(t)}
 {1-\exp(e^t)}]-\lambda\exp[-t-\exp(t)][1+\frac{\exp(t)}
 {1-\exp(e^t)}]^2G^{2}(y,t)
 \end{equation}

The semi-classical approximation of the Eq. (13)
\begin{equation}\partial_{y}z(y,t)\partial_{t}z(y,t)=c[1+\frac{\exp(t)}
{1-\exp(e^t)}]-\lambda\exp[-t-\exp(t)+z][1+\frac{\exp(t)}
{1-\exp(e^t)}]^2
\end{equation}
By using the method of characteristic, the solution of Eq. (14) are showed
in Fig. 2. \\

To conclude, recent experiments at HERA have supplied much information
about the nucleon structure at both large $Q^2$ and low $Q^2$. In
large $Q^2$, the DGLAP equation derived from perturbative QCD can describe
the behavior of parton distribution. The challenging problem is how to make
a unified treatment on nucleon structure at both large $Q^2$ and low $Q^2$.
This letter proposes a way to meet the requirement. It is well known that
the parton distribution
includes both perturbative QCD and nonperturbative QCD effects. The input
distribution reflects the nonperturbative QCD,
and the DGLAP equation itself is the perturbative QCD. So it seems that
the DGLAP equation describing the perturbative QCD effect and the input
distribution describing the nonperturbative QCD effect together can give the
comprehensive description to the parton distribution. However, it can be seen
clearly that the input distribution does not include all nonperturbative
effects, that is to say, some nonperturbative effects are reflected through
the running coupling. By considering the
nonperturbative effects in running coupling, it is a natural way to apply the
DGLAP equation in the low $Q^2$ region. So the evolution equation itself includes
both perturbative and nonperturbative effects. It should be noted that this way
is a work ansatz, which can not be derived from the theory. In viewing the Fig. 1,
it can be found that the nonperturbative contribution to DGLAP is very notable.

Although the predictions of GRV model by applying the DGLAP evolution
equation at very low $Q^2$ are in broad agreement with HERA experiments,
the evolution rate  which results from the model is faster than that of the
experiments.
By considering the nonperturbative contribution to DGLAP,
the discrepancy between GRV model and the experiments can be explained
naturally. From analysing the DGLAP equation, it can be
found that the running coupling determines the evolution rate,
which becomes slow by considering the nonperturbative contribution,
especially in very low $Q^2$ such as $Q^2=0.65GeV^2$.\\

Recently, one of the important questions is whether the shadowing effect
can be observed by the current experiments at HERA. Some people say ``can''
such as Shabelski and Treleani [15] while other people say ``cannot'' such as
Golec-Biernat, Krasny and Riess [16].
Ayala, Gay Ducati and Levin [17] argue that the shadowing effect is large in
the gluon distribution but small in $F_2(x,Q^2)$. Like the DGLAP equation,
the GLR equation can be treated by the same method.
In this paper, a firm
conclusion about this question does not be made. Nevertheless,
in viewing the Fig.2, it can be concluded that shadowing effect in the
GLR equation, which is modified by the nonperturbative contribution,
is not so notable as what has been studied
in the case when gluons  concentrate in "hot-spots" within proton
($R=2GeV^{-1}$).
By analysing the value of $\alpha_{s}(Q^{2})$ and $\alpha_{an}(Q^{2})$,
the simple explanation of this result is that the linear term in the GLR equation
is proportional to $\alpha_{s}(Q^{2})$ or $\alpha_{an}(Q^{2})$ while
the nonlinear term is proportional to $\alpha_{s}^{2}(Q^{2})$ or
$\alpha_{an}^{2}(Q^{2})$, so the shadowing effect is weakened
due to $\alpha_{s}(Q^2)>\alpha_{an}(Q^{2})$, especially at low $Q^2$ region.
The result means that the nonperturbative contribution weakens the shadowing
effect.
Some people [13] discussed the nuclear shadowing by applying the GLR
evolution equation without considering the nonperturbative contribution.
It might be interesting to restudy the nuclear shadowing by applying
the GLR equation (14), which is modified by the nonperturbative contribution.\\

In this paper, the DGLAP and GLR equation are solved by applying
the method of characteristic. From the initial conditions (12), it can be found
that the start point is $Q^2=4GeV^2$, which is not very low, because in very
low  $Q^2$ the evolution equation might be too complicated to treat and
the semi-classical approximation is not a good approximation.
However,  the conclusions shown in this paper can be deduced easily to the
very low $Q^2$ region, where the nonperturbative contribution becomes more
dominant  because  the difference between the formula (3)
and formula (5) $\alpha_{s}(Q^2)-\alpha_{an}(Q^2)$ which is $(0.5-0.4)=0.1$
at  $Q^2=0.65GeV^2$  is much more notable than that $(0.3-0.29)=0.01$ at
$Q^2=4GeV^2$.

In summary, it is believed that QCD, which includes
the perturbative part and the nonperturbative part, is a complete theory
to describe all strong interaction experiments. Nevertheless,
as the fundamental dynamical model, the DGLAP or GLR equation itself only
reflects the perturbative effect. Unfortunately, almost all strong
interaction experiments such as the deep-inelastic {\it ep} scattering
at HERA involve both
perturbative effect and nonperturbative effect. The purpose of the paper
is to develop a fundamental dynamical model which  itself includes
nonperturbative effect. By studying the model, we make the conclusions:
(i) the nonperturbative contribution suppresses
the evolution rate at the low $Q^2$, small-x region; (ii) the
nonperturbative contribution weakens the shadowing effect.
Those conclusions are helpful to explain the
recent experiments in low $Q^2$ region.

If the results of this paper are compared with the recent HERA data in detail,
the quark distribution must also be discussed in the low $Q^2$ region,  but it
should be noted that the recent HERA data is available at a few isolated values of
averaged x and $Q^2$, especially in very low $Q^2$, so how to analyse the HERA
experiments and compare the experiment data with the results of the model
developed in this paper will be a challenging work. Therefore in this
paper the quark distribution is not discussed.
However, the method developed in this paper can be easily extended to the evolution
equation for the quark distribution. Even the method can also be applied to
discuss the parton distribution at large-x, low $Q^2$ region, because
the DGLAP equation
modified by the nonperturbative contribution can be applied to both the large-x
region and small-x region (in small-x region, it is possible that the GLR equation
describes the parton behavior , but it is difficult to check the GLR equation
because the nonperturbative contribution weakens the shadowing effect showed
in this paper). \\

The main purpose of this paper is to developed a method, which the evolution
equation itself includes the nonperturbative contribution. The method has the
important theoretical feature that it suggests a smooth transition from the low $Q^2$ ("soft" ), where
nonperturbative contribution dominates, to the large $Q^2$ ("hard" ) region,
where the perturbative contribution dominates and the nonperturbative
contribution can be neglected.
Although the method is only a first step in considering the nonperturbative
contribution to QCD dynamical equations, it can be checked not only by the
HERA experiments but also by other strong interaction experiments because the
method proposed in this paper is adaptable to wide region.\\

The author  is grateful to C.Wang for his helpful discussions.
This work was supported in part by the Foundation of Shanghai Jiaotong University.\\

\noindent\bf {Figure Captions}\\

\rm Fig. 1. The dashed curve corresponds to the results of Eqs. (10)
and  the solid curve corresponds to the results of the DGLAP equation
without considering the nonperturbative contribution.\\

\rm Fig. 2. As for Fig. 1, except for the GLR equation in the case $R=2eV^2$.


\begin{thebibliography}{55}
\bibitem{1} L.N.Lipatov, Phys. Rep. 286 (1997) 131
\bibitem{2} B.Badelek, J.Kwiecinski, Rev. Mod. Phys. 68 (1996) 445
\bibitem{3} B.Badelek, K.Charchula, M.Krawczyk, and J.Kwiecinski, Rev. Mod.
Phys. 64(1992) 927
\bibitem{4} L.V.Gribov, E.M.Levin and M.G.Ryskin, Phys. Rep. 100 (1983) 1
\bibitem{5} E.M.Levin and M.G. Ryskin, Phys. Rep. 189 (1990) 269
\bibitem{6} I.Abt et al., Nucl. Phys. B 407 (1993) 515; M.Derrick et al.,
Phys. Lett. B 316 (1993) 412;
\bibitem{7} ZEUS Collaboration, M.Derrick et al., Phys. Lett. B 407 (1997) 432
\bibitem{8} M.Gl\"{u}ck, E.Reya and A.Vogt, Z. Phys. C 48 (1990) 471;
C 53 (1992) 127; C 67 (1995) 433
\bibitem{9} V.N.Gribov and L.N.Lipatov, Sov. J. Nucl. Phys. 15 (1972) 438, 675;
G.Altarelli and G.Parisi, Nucl. Phys. B 126 (1977) 298;
Yu.L.Dokshitzer, Sov. Phys. JETP 46 (1977) 641;
L.N.Lipatov, Sov. J. Nucl. Phys. 20 (1975) 93;
\bibitem{10} L.V. Gribov, E.M. Levin and M.G. Ryskin, Nucl. Phys. B 188 (1981) 555;
A.H. Mueller and J. Qiu, Nucl. Phys. B268 (1986) 427
\bibitem{11} J.Kwiecinski, A.D. Martin and P.J. Sutton, Phys. Rev. D 44 (1991) 2640
\bibitem{12} J.Collins and J.Kwiecinski, Nucl. Phys. B 316 (1989) 307
\bibitem{13} K.J.Eskola, J.Qiu and X.N.Wang, Phys. Rev. Lett. 72 (1994) 36
\bibitem{14} D.V.Shirkov and I.L.Solovtsov, Phys. Rev. Lett. 79 (1997) 1209
\bibitem{15} Yu.M. Shabelski and D. Treleani, Phys. Lett. B 403 (1997) 364
\bibitem{16} K. Golec-Biernat, M.W. Krasny and S. Riess, Phys.Lett.B 33 (1994)
\bibitem{17} A.L. Ayala, M.B. Gay Ducati and E.M. Levin, Nucl. Phys. B 511 (1998) 355 \\
\end{thebibliography}
\end{document}